\documentstyle[12pt]{article}

\newlength{\dinwidth}
\newlength{\dinmargin}
\newcommand{\resection}[1]{\setcounter{equation}{0}\section{#1}}

\begin{document}
\vspace*{3cm}
\begin{center}
  \begin{Large}
  \begin{bf}
 LOW ENERGY EFFECTIVE LAGRANGIAN\\
 OF THE BESS MODEL\\
  \end{bf}
  \end{Large}
  \vspace{2cm}
  \begin{large}
L. Anichini$^{(a)}$, R. Casalbuoni$^{(a,b)}$ and S. De Curtis$^{(b)}$\\
  \end{large}
\vspace{4mm}
$(a)$ Dipart. di Fisica, Univ. di Firenze, Largo E. Fermi, 2 - 50125
Firenze (Italy)\\
$(b)$ I.N.F.N., Sezione di Firenze, Largo E. Fermi, 2 - 50125
Firenze (Italy)\\
\end{center}
  \vspace{2.5cm}
\begin{center}
  \begin{bf}
  ABSTRACT
  \end{bf}
\end{center}
  \vspace{5mm}
\noindent
In this paper the low energy limit of the BESS model is studied in a
systematic way. The method consists in eliminating the heavy vector field,
by use of its classical equations of motion, in the infinite mass limit.
After the elimination of the heavy degrees of freedom we get additional
terms to the Standard Model lagrangian. After a finite renormalization
of the ordinary gauge bosons wave functions, and redefinition of the
lagrangian couplings in terms of $M_Z$, the fine structure constant and the
Fermi constant, we can read directly the deviations from the Standard Model.
By this procedure we can extend a result previously derived to the case in
which the heavy vector bosons have a direct coupling to fermions.
Consequences for the anomalous trilinear couplings are discussed.

\vspace{1.5cm}
\begin{center}
Firenze Preprint - DFF-210/10/1994
\end{center}

\newpage
\newcommand{\f}[2]{\frac{#1}{#2}}
\def\lq{\left [}
\def\rq{\right ]}
\def\dmu{\partial_\mu}
\def\dnu{\partial_\nu}
\def\dmus{\partial^\mu}
\def\dnus{\partial^\nu}
\def\gp{g'}
\def\gpt{{\tilde g'}}
\def\gs{g''}
\def\ggs{\frac{g}{\gs}}
\def\eps{{\epsilon}}
\def\tr{{\rm {tr}}}
\def\V{{\bf{V}}}
\def\W{{\bf{W}}}
\def\Wt{\tilde{\bf {W}}}
\def\Y{{\bf{Y}}}
\def\Yt{\tilde{\bf {Y}}}
\def\L{{\cal L}}
\def\s{s_\theta}
\def\st{\tilde s_\theta}
\def\c{c_\theta}
\def\ct{\tilde c_\theta}
\def\gt{\tilde g}
\def\et{\tilde e}
\def\At{\tilde A}
\def\Zt{\tilde Z}
\def\Wpt{\tilde W^+}
\def\Wmt{\tilde W^-}
\newcommand{\be}{\begin{equation}}
\newcommand{\ee}{\end{equation}}
\newcommand{\bea}{\begin{eqnarray}}
\newcommand{\eea}{\end{eqnarray}}
\newcommand{\nn}{\nonumber}
\newcommand{\dd}{\displaystyle}
\setcounter{page}{1}
\resection{Introduction}

The experimental data on electroweak interactions are giving more and more
confirmations about the validity of the electroweak Standard Model (SM)
up to energies of roughly 100 $GeV$. One of the key ingredients of the SM
is the electroweak symmetry breaking, for which, however, the theory does not
provide an adequate mechanism.
In fact, the minimal SM with a single Higgs field, given its well known
patologies, can be considered, at most, a good parametrization of the symmetry
breaking.

The solutions to this problem proposed so far can be divided into two broad
classes depending if the Higgs is considered elementary or composite.
Supersymmetric models  belong to the first class, because
supersymmetry gives rise naturally to the cancellation of the quadratic
divergences in the Higgs self-mass, avoiding the fine tuning  problem.
To the second class belong all the models in which a new strong interaction
theory is required at a scale of about $1~TeV$. The prototype of these models
is technicolor \cite{technicolor}.
Whereas the models of the first class are weak interacting and therefore
perturbatively calculable, in the other case life is more difficult.
In fact, most of the calculations made in this area are based on the scaling
from usual strong interactions.
Other possibilities of dealing with the strong interacting sector, rely on the
chiral perturbation theory.
In fact the idea of technicolor is that the new strong interaction provides
the dynamical breaking of $SU(2)_L \otimes U(1)_Y$. Then, one can use the
chiral
approach to write down an effective Lagrangian describing the Goldstone bosons
arising from the symmetry breaking and which represent the longitudinal degrees
of the $W$ and $Z$ mesons \cite{cpt}.
 Typically this expansion is arrested to the fourth
order and one extrapolates to higher energies using some
unitarization procedures \cite{dob}.

Another possibility is to introduce in the effective theory, in addition to the
Goldstone bosons, other resonances due to the strong interaction. Such
possibility was considered several years ago in the context of the BESS model
(BESS stands for Breaking Electroweak Symmetry Breaking)
\cite{bess}, where, on the basis of chiral invariance, vector resonances are
introduced.
The phenomenology of this model has been studied in various papers as well as
the limitations on its parameters arising from weak-interaction experiments.
 Regarding this last point, the studies have been done mainly
numerically
\cite{besslimit} or, in special case, analitically \cite{self}.

In this paper we want to study the problem of the experimental limitations on
the BESS parameter space in a more general way. This treatment will give us the
possibility to generalize it to the model which includes also
axial-vector resonances, that have been already proposed
in ref. \cite{assiali}.

The idea is very simple: we eliminate the fields of the vector resonances from
the Lagrangian via their classical equation of motions in the limit of infinite
mass, which in physical terms means that the mass must be much bigger than
$M_W$. By defining in a convenient way this limit, we get an effective
Lagrangian in terms of $W,~Z$ and the photon fields, from which after a finite
renormalization \cite{burges} we can read directly the deviations from the SM.

In Sect. 2 we review the basic ingredients of the BESS model. In Sect. 3  we
perform the low-energy limit by considering $M_V\to \infty$
(limit that will be conveniently defined later on), $V$  being the new
resonances. In Sect. 4 we identify the physical quantities by performing
fields and couplings renormalization. The results are in Sect. 5 and 6.
The low-energy effects of the $V$ particles are discussed in Sect. 5 in
terms of the $\epsilon$ parameters, and bounds on the BESS parameter space
from the most recent experimental measurements are derived.
In Sect. 6 we calculate the corrections to the trilinear
gauge boson couplings which will be of great phenomenological
interest for the studies  at the future $e^+e^-$ colliders. For completeness
we give also the expressions for the anomalous quadrilinear couplings.
Finally we discuss the results in Sect. 7.

\resection{The BESS model}

An effective description of the symmetry breaking mechanism in electroweak
theories can be done in
terms of a non linear $\sigma$-model formulated on the
quotient space of the breaking of $SU(2)_L\otimes SU(2)_R \to SU(2)_{L+R}$.
This is the case when considering the limit of strong interacting Higgs sector
($M_H\to\infty$).

As it is known \cite{bala}, this non linear $\sigma$-model
possesses a hidden local symmetry $H_{local}=SU(2)_V$.
Our assumption is that the
appearing of this symmetry is realized through a new triplet of dynamical
vector boson resonances $\V$ \cite{bess}.

The main steps for the construction of the BESS model lagrangian are
the following.
One introduces the group variables $g(x)\in G=SU(2)_L\otimes SU(2)_R$
\be
g(x)=(L(x),R(x))
\ee
with $L(x)\in SU(2)_L$ and $R(x)\in SU(2)_R$ which transform under
$G\otimes H_{local}$ group as follows: $L\to g_L L h(x)$, $R\to g_R R h(x)$
with $g_{L,R}\in SU(2)_{L,R}$ and $h(x)\in SU(2)_{V}$.

One further introduces the Maurer-Cartan form
\be
\omega_\mu= g^\dagger\dmu g=(L^\dagger\dmu L,R^\dagger\dmu R)
\ee
which can be decomposed into $\omega_\mu^\parallel$ lying in the Lie algebra
of $H_{local}$, and into the orthogonal complement $\omega_\mu^\perp$
\bea
\omega^{\parallel}_\mu &=& \f{1}{2} (L^\dagger\dmu L+R^\dagger\dmu R)\nn\\
\omega^{\perp}_\mu &=& \f{1}{2} (L^\dagger\dmu L-R^\dagger\dmu R)
\eea
Both $\omega_\mu^\parallel$ and $\omega_\mu^\perp$ are singlet of $G$
and transform under $H_{local}$ as
\bea
\omega^{\parallel}_\mu &\to & h^\dagger\omega^{\parallel}_\mu h+
h^\dagger\dmu h\nn\\
\omega^{\perp}_\mu &\to & h^\dagger\omega^{\perp}_\mu h
\eea
The non linear $\sigma$-model Lagrangian describing the electroweak symmetry
breaking sector can be easily reconstructed in terms
of $\omega^{\perp}_\mu$
\be
{\cal L} =
-v^2\tr (\omega^{\perp}_\mu\omega^{\perp\mu})=\f{v^2}{4}\tr (\dmu U\dmus
U^\dagger)
\ee
where $U=LR^\dagger$ is a singlet under $H_{local}$ and $v\simeq 250~GeV$
is the standard electroweak scale.

Introducing a triplet of gauge bosons $\V_\mu$ for the local group $SU(2)_V$,
one can show that the most general Lagrangian, symmetric under
$G\otimes H_{local}$ and under the parity transformation $L\leftrightarrow R$,
containing at most two derivatives,
can be constructed as an arbitrary combination of two invariant terms.
Furthermore, assuming that the gauge bosons of the hidden symmetry
become dynamical we get

\bea
{\cal L} &=&
-v^2\Big[\tr (\omega^{\perp}_\mu\omega^{\perp\mu})
 +\alpha~ \tr (\omega^{\parallel}_\mu-\V_\mu)^2 \Big]\nn\\
& &+\f{2}{\gs^2}\tr[F^{\mu\nu}(\V)F_{\mu\nu}(\V)]\eea
with $\alpha$ an arbitrary parameter,
\be
F_{\mu\nu}(\V)  = \dmu \V_\nu-\dnu \V_\mu+[\V_\mu,\V_\nu]
\ee
and $\V_\mu=\f{i}{2}\f{\gs}{2} V_\mu ^a\tau^a$, with $\gs$ the new gauge
coupling constant and $\tau^a$ the Pauli matrices.

The gauging of the standard $SU(2)_L\otimes U(1)_Y$ group is simply obtained
by substituting in (2.3) the ordinary derivatives with covariant left and
right derivatives acting on the left and right group elements respectively
\bea
\dmu L &\to & (\dmu +\Wt_\mu) L\nn\\
\dmu R &\to & (\dmu +\Yt_\mu) R
\eea
where $\Wt_\mu=\f{i}{2}\gt {\tilde W}_\mu ^a\tau^a$ and
$\Yt_\mu=\f{i}{2}\gpt {\tilde Y}_\mu\tau^3$ and by
adding the standard kinetic terms for $\Wt$ and $\Yt$
\be
{\cal L}^{kin}(\Wt,\Yt) =
\f{1}{2 \gt^2}\tr[F^{\mu\nu}(\Wt)F_{\mu\nu}(\Wt)]
   +\f{1}{2 {\tilde g}^{\prime 2}}\tr[F^{\mu\nu}(\Yt)F_{\mu\nu}(\Yt)]
\ee
with
\bea
F_{\mu\nu}(\Wt)  &=& \dmu \Wt_\nu-\dnu \Wt_\mu+[\Wt_\mu,\Wt_\nu]\nn\\
F_{\mu\nu}(\Yt)  &=& \dmu \Yt_\nu-\dnu \Yt_\mu
\eea

Of course, if one assumes only the invariance under $SU(2)_L\otimes U(1)_Y
\otimes SU(2)_V$ many other invariant terms are possible, but, for simplicity,
we will ignore these extra terms in this paper.

Due  to the gauge invariance of $\L$ we can choose the gauge with
$L=R=1$ \cite{bess}(unitary gauge) and we get

\be
\L=-\f{v^2}{4}\Big[\tr(\Wt_\mu-\Yt_\mu)^2
+\alpha \tr(\Wt_\mu+\Yt_\mu-2\V_\mu)^2\Big]
+\L^{kin}(\Wt,\Yt,\V)
\ee
We have used tilded quantities to remember that, due to the effects
of the $\V$ particles, they are not the physical parameters and fields.
In the next sections we will derive the relations between the tilded
quantities and the physical ones in the low-energy limit.

{}From eq. (2.11) one can easily derive the mass eigenstates and the mixing
angles among the standard gauge bosons and the new resonances \cite{bess}.
Furthermore, since in the limit $\gs\to\infty$, the lagrangian $\L$
reproduces the SM terms, corrections to the SM relations come in powers
of $1/\gs$.

Finally let us consider the fermions of the SM and denote them by $\psi_L$
and $\psi_R$. They couple to $\V$ via the mixing with the standard $\Wt$
and $\Yt$.
In addition, we also expect direct couplings to the new vector bosons
since they are allowed by the symmetries of $\L$ \cite{bess}.
In fact, we can define Fermi fields transforming as doublets under the
local group $SU(2)_V$ and singlets under the global one: $\chi_L=L^\dagger
\psi_L$. We can then construct an invariant term acting on $\chi_L$ by
the covariant derivative with respect to $SU(2)_V$.
In the unitary gauge we get
\bea
\L_{fermion} &=& \overline{\tilde \psi}_L i \gamma^\mu\Big(\dmu+\Wt_\mu+
                      \f{i}{2}\gpt(B-L){\tilde Y}_\mu\Big){\tilde \psi}_L\nn\\
     &+&\overline{\tilde \psi}_R i \gamma^\mu\Big(\dmu+\Yt_\mu+\f{i}{2}\gpt
                      (B-L){\tilde Y}_\mu\Big){\tilde \psi}_R\nn\\
     &+& b \overline{\tilde \psi}_L i \gamma^\mu\Big(\dmu+\V_\mu+
                      \f{i}{2}\gpt(B-L){\tilde Y}_\mu\Big){\tilde \psi}_L
\eea
where $B(L)$ is the baryon (lepton) number and
$b$ is a new parameter. Notice that
due to the introduction of the direct coupling of the $\V$ to the
fermions, we have to rescale ${\tilde \psi}_L=
 (1+b)^{-1/2}\psi_L$ in order to get
a canonical kinetic term for the fermions \cite{bess}.

\resection{The low-energy limit}

We want to study the effects of the $\V$ particles in the low-energy limit.
This can be done by eliminating the $\V$ fields with the solution
of their equations of motion for $M_V\to\infty$. In fact in this
limit the kinetic term of the new resonances is negligible.
Also, since their  mass is given by (neglecting electromagnetic corrections)
$M_V^2\simeq\alpha (v^2/4) \gs^2$, we will take the limit by fixing $\gs$
and $v$ and sending $\alpha\to\infty$.
Because $v$ is experimentally fixed, the only other
possibility would be to send $\gs\to\infty$, but, in this case, the $\V$ bosons
would decouple.

We will study the
effective theory by considering corrections up to order $(1/\gs)^2$.

Concerning the other parameter $b$, we expect it to be of the order
$(1/\gs)^2$ as arising from $\V-{\bf W}$ mixing and one-loop corrections to the
$\bf W$-fermion vertex and $\bf W$ propagator \cite{bess}. The present
bounds from the LEP measurements, are consistent with this conjecture
(see Sect. 5)
so we will consider our effective theory as an expansion both in
$1/\gs$ and $b$ in which terms of the order $b/\gs$ are neglected.

Let us solve the equation of motion for $\V$ in this limit. By evaluating
$\partial\L/\partial V^a_\mu$ from eqs. (2.11) and (2.12) we get
\be
\V_\mu=\f{1}{2}\Big(\Wt_\mu+\Yt_\mu\Big)
\ee
By substituting this equation in the Lagrangian we get
\bea
\L_{eff} &=& -\f{v^2}{4}\tr(\Wt_\mu-\Yt_\mu)^2\nn\\
    &+& \f{2}{\gs^2}\tr\Big[F^{\mu\nu}\Big(\f{\Wt+\Yt}{2}\Big)
                         F_{\mu\nu}\Big(\f{\Wt+\Yt}{2}\Big)\Big]\nn\\
    &+& \L^{kin}(\Wt,\Yt)+\L_{eff}^{charged}+\L_{eff}^{neutral}+\overline\psi
          i\gamma^\mu\dmu\psi
\eea
where we have separated the charged
 and neutral fermionic sector
\bea
\L_{eff}^{charged} &=& -\f{\et}{\sqrt{2} \st}
\big(1-\f{b}{2}\big)\overline\psi_d
     \gamma^\mu\f{1-\gamma_5}{2}\psi_u \Wmt_\mu +~h.c.\\
\L_{eff}^{neutral} &=& -\f{\et}{\st \ct} \big(1-\f{b}{2}\big)\overline\psi
     \gamma^\mu\Big[ T^3_L \f{1-\gamma_5}{2}-Q \st^2 \big(1+\f{b}{2}\big)\Big]
       \psi \Zt_\mu - \et \overline\psi \gamma^\mu Q \psi \At_\mu
\eea
and we have used the following standard definitions:
\bea
Q &=& \f{\tau^3}{2}+\f{B-L}{2}\nn\\
T^3_L \psi_L &=& \f{\tau^3}{2}\psi_L~~~~~~T^3_L \psi_R =0\nn\\
{\tilde W}_\mu^\pm &=&\f{1}{\sqrt{2}}({\tilde W}_1\mp i {\tilde W}_2)\nn\\
{\tilde W}_\mu^3 &=& \st \At_\mu+\ct \Zt_\mu\nn\\
{\tilde Y}_\mu &=& \ct \At_\mu-\st \Zt_\mu\nn\\
\et &=& \gt\st = \gpt\ct\nn\\
\psi &=&\left(
\begin{array}{c}
\psi_u\\
\psi_d\end{array}
\right)
\eea

Notice that $\gt,\gpt,\et,\st,\ct$ have the same definitions
 as in the SM. As stated before,
 due to the effects of the $\V$ particles, these
are not the physical quantities in our model.

{}From eq. (3.2) we see that the effective contribution of the $\V$ particles
give additional terms to the kinetic terms of the standard $\Wt$ and $\Yt$.
This will imply a renormalization of the fields in order
to have canonical kinetic terms.
By calling  $\L^{kin~(2)}_{eff}$ the bilinear terms coming from the kinetic
terms for the gauge bosons, we have (using the definitions of eq. (3.5))
\bea
\L^{kin~(2)}_{eff}({\tilde W}^\pm,\At,\Zt) &=&
 -\f{1}{4} (1+z_\gamma)\At_{\mu\nu}\At^{\mu\nu}
-\f{1}{2} (1+z_w){\tilde W}_{\mu\nu}^+ {\tilde W}^{\mu\nu-} \nn\\
&-& \f{1}{4} (1+z_z)\Zt_{\mu\nu}\Zt^{\mu\nu}+\f{1}{2} z_{z\gamma}
\At_{\mu\nu}\Zt^{\mu\nu}
\eea
where $O_{\mu\nu}=\dmu O_\nu-\dnu O_\mu$, ($O={\tilde W}^\pm,\At,\Zt$) and
\be
z_\gamma = 4 \s^2\Big(\f{g}{\gs}\Big)^2~~~~~
z_w =  \Big(\f{g}{\gs}\Big)^2~~~~~
z_z =   \f{c^2_{2\theta}}{\c^2}\Big(\f{g}{\gs}\Big)^2~~~~~
z_{z\gamma} =  -2 \f{\s}{\c}c^2_{2\theta}  \Big(\f{g}{\gs}\Big)^2
\ee
In eq. (3.7) we have not used the tilded quantities since these parameters
are already of the order of $(1/\gs)^2$.

The corrections to $\L_{SM}$ are $U(1)_{em}$ invariant and produce
a wave-function renormalization of $\At_\mu,\Zt_\mu,{\tilde W}_\mu^\pm$ plus a
mixing term $\At_\mu-\Zt_\mu$. Notice that in general there could be two other
renormalization terms: $\delta M_W^2 \Wpt_\mu{\tilde W}^{\mu-}$
and  $\delta M_Z^2 \Zt_\mu\Zt^\mu$
which, however, are zero in this model.
In the next section we will absorb these corrections
by a convenient redefinition of the fields. Actually there are only
three renormalization transformations of the fields
$\At_\mu,\Zt_\mu,{\tilde W}_\mu^\pm$ without changing the physics. This
means that three of the four deviations $z_\gamma,z_w,z_z,z_{z\gamma}$ are
non physical, and this is consistent with the fact that they depend on a
single parameter $\gs$.

Analogously we can calculate the corrections to the trilinear and
quadrilinear terms coming from the kinetic terms of the gauge bosons. We get
\bea
\L^{kin~(3)}_{eff}({\tilde W}^\pm,\At,\Zt) &=&
i\gt\ct(1+z_{zww})\Big[\Zt^{\mu\nu}\Wmt_\mu\Wpt_\nu +
       \Zt^{\nu}({\tilde W}^-_{\mu\nu}\Wpt_\mu-
        {\tilde W}^+_{\mu\nu}\Wmt_\mu)\Big]\nn\\
& &+i\et(1+z_w)\Big[\At^{\mu\nu}\Wmt_\mu\Wpt_\nu +
       \At^{\nu}({\tilde W}^-_{\mu\nu}\Wpt_\mu-
        {\tilde W}^+_{\mu\nu}\Wmt_\mu)\Big]
\eea
\bea
\L^{kin~(4)}_{eff}({\tilde W}^\pm,\At,\Zt) &=&
S_{\mu\rho\nu\sigma}\Wpt_\mu\Wmt_\rho\Big[-\f{\et^2}{2}(1+z_w)
   \At_\nu\At_\sigma-\et\gt\ct(1+z_{zww})\At_\nu\Zt_\sigma\nn\\
& &+\f{1}{2}\gt^2(1+z_{wwww})\Wpt_\nu
  \Wmt_\sigma-\f{1}{2}\gt^2\ct^2(1+z_{zzww})\Zt_\nu\Zt_\sigma\Big]
\eea
where $S_{\mu\rho\nu\sigma}=2g_{\mu\rho}g_{\nu\sigma}-
g_{\mu\nu}g_{\rho\sigma}-g_{\mu\sigma}g_{\rho\nu}$ and
\be
z_{zww}=\f{c_{2\theta}}{2\c^2}\Big(\f{g}{\gs}\Big)^2~~~~~
z_{wwww}=\f{1}{4}\Big(\f{g}{\gs}\Big)^2~~~~~
z_{zzww}=\f{c^2_{2\theta}}{4\c^4}\Big(\f{g}{\gs}\Big)^2
\ee
The electromagnetic $U(1)$ invariance is preserved.

\resection{Fields and couplings renormalization}

To identify the physical quantities we define new fields in such a way
to have canonical kinetic terms and to cancel the mixing term
$\At_\mu-\Zt_\mu$. They are the following:
\bea
\At_\mu &=& (1-\f{z_\gamma}{2}) A_\mu+z_{z_\gamma} Z_\mu\nn\\
{\tilde W}^\pm_\mu &=& (1-\f{z_w}{2}) W_\mu^\pm\nn\\
\Zt_\mu &=& (1-\f{z_z}{2}) Z_\mu
\eea

Let us study the effects of this renormalization.

First of all for the
mass terms we get:
\be
-\f{v^2}{4}\tr(\Wt_\mu-\Yt_\mu)^2=-{\tilde M}_W^2 (1-z_w) W^{\mu+}W^-_\mu
-\f{1}{2}{\tilde M}_Z^2 (1-z_z) Z^\mu Z_\mu
\ee
with
\be
{\tilde M}_W^2 =\f{v^2}{4}\gt^2 ~~~~~~~~{\tilde M}_Z^2={\tilde M}_W^2/\ct^2
\ee

Also, the field renormalization affects all the couplings of the standard
gauge bosons to the fermions. By substituting eq. (4.1)
in eqs. (3.3) and (3.4)
we get
\bea
\L_{eff}^{charged} &=& -\f{\et}{\sqrt{2} \st} \big(1-\f{b}{2}-\f{z_w}{2}\big)
     \overline\psi_d
     \gamma^\mu\f{1-\gamma_5}{2}\psi_u W^-_\mu +~h.c.\\
\L_{eff}^{neutral} &=& -\f{\et}{\st \ct} \big(1-\f{b}{2}-\f{z_z}{2}\big)
      \overline\psi
     \gamma^\mu\Big[ T^3_L \f{1-\gamma_5}{2}-Q \st^2 \big(1+\f{b}{2}
      -\f{\ct}{\st}z_{z \gamma}\big)\Big]
       \psi Z_\mu\nn\\
     & &  - \et \big(1-\f{z_\gamma}{2}\big)
        \overline\psi \gamma^\mu Q \psi A_\mu
\eea

We see that the physical constants as the electric charge, the Fermi constant
and the mass of the $Z$, which are the input parameters for
 the physics at LEP, must be redefined in terms of the parameters appearing
in our effective lagrangian. They are identified as follows
\bea
e &=& \et \big(1-\f{z_\gamma}{2}\big)\nn\\
M_Z^2 &=& {\tilde M}_Z^2 (1-z_z)
\eea
Concerning the Fermi constant $G_F$, it is evaluated from the $\mu$-decay
process. Since the charged current coupling is modified by a factor
$(1-b/2-z_w/2)$ and the $W$ mass
\be
M_W^2 = {\tilde M}_W^2 (1-z_w)
\ee
 we get
\be
\f{G_F}{\sqrt{2}}=\f{\et^2(1-b-z_w)}{8\st^2{\tilde M}_W^2(1-z_w)}
     =\f{e^2}{8\st^2\ct^2 M^2_Z} (1-b-z_z+z_\gamma)
\ee
where we have used eq. (4.6).
We choose to define $\s$ and $\c$ by equating this expression to the one
in the SM (tree level):
$G_F/\sqrt{2}=e^2/(8\s^2\c^2 M^2_Z)$. We get
\be
\s^2\c^2=\st^2\ct^2(1+b+z_z-z_\gamma)
\ee
that is
\bea
\s^2 &=& \st^2 \big( 1+\f{\c^2}{c_{2\theta}}(b+z_z-z_\gamma)\big)\nn\\
\c^2 &=& \ct^2 \big( 1-\f{\s^2}{c_{2\theta}}(b+z_z-z_\gamma)\big)
\eea

\resection{Observables}

Let us now discuss how the effects of the $\V$ modifies the observables.

For the physics at LEP and TEVATRON, the modifications due to heavy particles
are contained in the so-called oblique corrections. In the low-energy limit,
one can expand the vacuum polarization amplitudes in $q^2/M^2_V$ and they
can be parametrized in terms of three independent parameters.
They are for example $\Delta r_W$, $\Delta k$ and $\Delta\rho$
or, equivalently, the $\eps$ parameters \cite{altarelli}.

Let us start from $\Delta r_W$. It is customary define
\be
\f{M^2_W}{M^2_Z}=
 \c^2\Big[1-\f{\s^2}{c_{2\theta}}\Delta r_W\Big]
\ee
{}From the relation ${\tilde M}^2_W=
{\tilde M}^2_Z\ct^2$ we get
\be
\f{M^2_W}{M^2_Z}=
 \c^2\Big[1+ z_z-z_w-\f{\s^2}{c_{2\theta}}(-b+z_\gamma-z_z)\Big]
\ee
so, for comparison, and using eq. (3.7)
\be
\Delta r_W=-b+z_\gamma+\f{c_{2\theta}}{\s^2}z_w-\f{\c^2}{\s^2}z_z=-b+2
\Big(\f{g}{\gs}\Big)^2
\ee

The definitions of $\Delta\rho$ and $\Delta k$
are given in terms of the neutral current
coupling to the $Z$
\be
\L^{neutral}(Z)=-\f{e}{\s\c}\Big(1+\f{\Delta\rho}{2}\Big)Z_\mu\overline\psi
[\gamma^\mu g_V+\gamma^\mu \gamma_5g_A]\psi
\ee
with
\bea
g_V &=& \f{T^3_L}{2}-s^2_{\bar\theta} Q\nn\\
g_A &=& -\f{T^3_L}{2}\nn\\
s^2_{\bar\theta} &=& (1+\Delta k) \s^2
\eea

By using eqs. (4.6) and (4.10) we get
\be
\f{e}{\s\c}=\f{\et}{\st\ct}\Big(1-\f{b}{2}-\f{z_z}{2}\Big)
\ee
For comparison with eq. (4.5), and using eq. (3.7), we obtain
\bea
\Delta\rho &=& 0\nn\\
\Delta k &=& \f{\c^2}{c_{2\theta}}(z_\gamma-z_z)-\f{\c}{\s}z_{z\gamma}
-\f{1}{2 c_{2\theta}} b=\f{1}{c_{2\theta}}\Big[-\f{b}{2}
+\Big(\f{g}{\gs}\Big)^2\Big]
\eea

Summarizing, we have the following correspondence between corrections and
observables: $\Delta r_W$ is equivalent to $M_W/M_Z$ which is
measured at TEVATRON, $\Delta k$ modifies the
vector coupling $g_V$ and $\Delta\rho$ modifies the neutral coupling
overall strength.
At LEP, $\Delta k$ can be obtained by measuring the forward-backward
 asymmetry at the $Z$
peak. Then, having fixed $\Delta k$,
 $\Delta\rho$ can be determined by the leptonic width.
All these quantities receive contributions also from weak radiative
corrections. In particular they depend quadratically from the top mass
which is still affected by a large error.
{}From the point of view of data analysis it turns out to be more convenient
to isolate such contribution in $\Delta\rho$ and define two other linear
combinations which depend only logarithmically on $m_{top}$. They are
the so-called $\eps$ parameters \cite{altarelli}
\bea
\eps_1 &=& \Delta\rho\nn\\
\eps_2 &=& \c^2\Delta\rho+\f{\s^2}{c_{2\theta}}\Delta r_W-2 \s^2\Delta k\nn\\
\eps_3 &=& \c^2\Delta\rho+c_{2\theta}\Delta k
\eea
Using  eqs. (5.3) and (5.7) we get
\bea
\eps_1 &=& 0\nn\\
\eps_2 &=& 0\nn\\
\eps_3 &=& -\f{b}{2}+\Big(\f{g}{\gs}\Big)^2
\eea
This extends the previous results obtained in \cite{self} to the case $b\ne 0$.

We can derive restrictions on the BESS parameters by using the experimental
data on $\eps_3$.
The most recent value for $\eps_3$ obtained by combining the LEP, low-energies,
CDF/UA2 and SLD data \cite{dati} gives
\be
\eps_3=(3.9\pm 1.7)\times 10^{-3}
\ee
By assuming for the BESS model the same one-loop radiative corrections
as for the SM in which the Higgs mass is used as a cut-off $\Lambda$
\cite{besslimit}, we add to $\eps_3$ given in eq. (5.9) the contribution coming
from the radiative corrections \cite{dati}
calculated for $M_H=\Lambda=1~TeV$ and $m_{top}=(174\pm 17)~GeV$, which are
$(\eps_3)_{rad.corr.}\simeq (6.39_{+0.20}^{-0.14})\times 10^{-3}$.
The allowed region at $90\%$ C.L. in the plane $(b,g/\gs)$ is
shown in Fig. 1. By increasing the value of the top mass, the region moves
slightly to the left (the solid (dashed) line is for $m_t=191(157)~GeV$).

For completeness, we derive the correction to the charged current
coupling which is defined as follows
\be
\L^{charged} = -h_W\overline\psi_d
     \gamma^\mu\f{1-\gamma_5}{2}\psi_u W^-_\mu +~h.c.
\ee
where, in the SM, $h_W=e/(\sqrt {2}\s)$.
By comparing with eq. (4.4) and using eqs. (4.6) and (4.10) we get
\be
h_W=\f{e}{\sqrt{2}\s}\Big[1+\f{z_\gamma}{2}-\f{z_w}{2}-\f{b}{2}
+\f{\c^2}{2c_{2\theta}}(z_z-z_\gamma+b)\Big]=
\f{e}{\sqrt{2}\s}\Big[1-\f{\s^2}{c_{2\theta}}\eps_3\Big]
\ee
where we have used the eqs. (3.7) and (5.9).

It is worth to stress that, in the low-energy regime (that is up to the mass of
the $Z$),
 all deviations to the Standard Model due to BESS are contained in
$\epsilon_3$, which turns out to be the sum of two contributions, one from
the direct coupling of ${\bf V}$ to fermions, and the other coming out from the
mixing of ${\bf V}$ with the ordinary gauge bosons. As a consequence the low
energy experiments restrict the plane $(b,g/\gs)$ only to a strip
(see Fig. 1), and not to a closed region. However, more informations can be
obtained by studying the trilinear couplings.

\resection{Anomalous trilinear and quadrilinear gauge couplings}

The corrections to the trilinear and quadrilinear gauge couplings
come from three different sources:
from the $\V$ kinetic term, after the substitution of the
equation of motion for the $\V$ fields, from the $\W$ and $\Y$
kinetic terms after renormalization, from the renormalization
of the couplings.

Let us perform the renormalization of the fields and couplings as
defined in Sect. 4 to evaluate the anomalous contributions to the
trilinear and quadrilinear gauge couplings given in eqs. (3.8) and (3.9).
The result is the following
\bea
\L^{kin~(3)}_{eff}( W^\pm,A,Z) &=&
 i g \c (1+k)\Big[Z^{\mu\nu}W^-_\mu W^+_\nu +
       Z^{\nu}(W^-_{\mu\nu}W^+_\mu-
        W^+_{\mu\nu}W^-_\mu)\Big]\nn\\
& &+i e\Big[A^{\mu\nu}W^-_\mu W^+_\nu +
       A^{\nu}( W^-_{\mu\nu} W^+_\mu-
        W^+_{\mu\nu}W^-_\mu)\Big]
\eea

\bea
\L^{kin~(4)}_{eff}(W^\pm,A,Z) &=&
S_{\mu\rho\nu\sigma}W^+_\mu W^-_\rho\Big[-\f{e^2}{2}
   A_\nu A_\sigma-e g\c (1+k)A_\nu Z_\sigma\nn\\
& &+\f{1}{2} g^2(1+k_1)W^+_\nu
  W^-_\sigma-\f{1}{2} g^2\c^2(1+k_2)Z_\nu Z_\sigma\Big]
\eea
with
\bea
k &=& \f{1}{2 c_{2\theta}} b -
\f{1}{2\c^2 c_{2\theta}}\Big(\f{g}{\gs}\Big)^2 \nn\\
k_1 &=&\f{\c^2}{c_{2\theta}} b+\Big(\f{1}{4}-
\f{1}{c_{2\theta}}\Big)\Big(\f{g}{\gs}\Big)^2\nn\\
k_2 &=& \f{1}{c_{2\theta}} b-
\f{1+2\c^2}{4\c^4 c_{2\theta}}\Big(\f{g}{\gs}\Big)^2
\eea

If we want to write an effective Lagrangian containing terms up to the
order $p^4$  (the gauge fields are formally considered of the order $p$)
then we have other invariant terms which can contribute to the anomalous
trilinear and quadrilinear couplings among the SM gauge bosons.
For example, two invariant terms which preserve $CP$ and $L\leftrightarrow R$
invariances, are the following (see the second paper in ref. \cite{bess})
\bea
L_I &=& \gamma ~\tr \Big(F^{\mu\nu} (\V) [\omega^{\perp}_\mu,
\omega^{\perp}_\nu]\Big)\nn\\
L_{II} &=& \delta ~\tr \Big(F^{\mu\nu} (\V) [\omega^{\parallel}_\mu-\V_\mu,
\omega^{\parallel}_\nu-\V_\nu]\Big)
\eea
Let us substitute the expressions (2.3). Then, after the $SU(2)_L
\otimes U(1)_Y$ gauging through eq. (2.8), we get,
in the unitary gauge
\bea
L_I &=& \f{\gamma}{4} \tr \Big(F^{\mu\nu} (\V) [\Wt_\mu-\Yt_\mu,
\Wt_\nu-\Yt_\nu]\Big)\nn\\
L_{II} &=& \f{\delta}{4} \tr \Big(F^{\mu\nu} (\V) [\Wt_\mu+\Yt_\mu-2\V_\mu,
\Wt_\nu+\Yt_\nu-2\V_\nu]\Big)
\eea

Notice that if we add these two terms to the BESS Lagrangian (2.11), the
solution of the classical equations of motion for $\V$ are again given by eq.
(3.1), because the contributions from $L_I$ and $L_{II}$ vanish in the
limit $\alpha\to \infty$. Furthermore, when substituting the classical solution
(3.1), we see that only $L_{I}$ survives in the effective Lagrangian.
By performing the explicit calculations and separating the trilinear from the
quadrilinear terms, we get

\bea
{(L_{I})}^{(3)}_{eff}( W^\pm,A,Z) &=&
 i \gamma \f{g^2}{4}\Big [
g \f{c_{2\theta}}{2  c_\theta}
Z^{\mu\nu}W^-_\mu W^+_\nu+
     \f{g}{2 c_\theta}
  Z^{\nu}(W^-_{\mu\nu}W^+_\mu-
        W^+_{\mu\nu}W^-_\mu)\nn\\
 & &+e A^{\mu\nu}W^-_\mu W^+_\nu\Big]
\eea

\bea
{(L_{I})}^{(4)}_{eff}(W^\pm,A,Z) &=& \gamma\f{g^2}{4}
S_{\mu\rho\nu\sigma}W^+_\mu W^-_\rho\Big[
-e \f{g}{2 \c} A_\nu Z_\sigma\nn\\
& &-\f{g^2}{4} W^+_\nu
  W^-_\sigma- g^2\f{c_{2\theta}}{4 c^2_\theta}
Z_\nu Z_\sigma\Big]
\eea

Notice that since we want to recover the SM Lagrangian
in the $\gs\to\infty$ limit, we consider $\gamma$ of the
order of $1/\gs$.
In this way, it is not necessary to perform the
renormalization of the fields and couplings as described in Sect. 4 since
all those corrections give contributions at least of the order $(1/\gs)^3$
which are negligible in our context.

Concerning the anomalous trilinear terms, we end up with the following
observation. By comparing with ref. \cite{bil} we can extract relations
among our parameters $k$ and $\gamma$ and the definitions used in the
literature to parametrize the deviations from the SM trilinear couplings
to be tested at the future colliders. We get
\be
x_\gamma = \gamma \f{g^2}{4}~~~~~
\delta_z = \f{\c}{\s}\Big(k+\gamma \f{g^2}{8 c^2_\theta}\Big)~~~~~
x_z = -\gamma \f{g^2}{4}\f{\s}{\c}
\ee
where $\delta_z$ describes a deviation of the $ZW^+W^-$ overall coupling from
the standard value while $x_\gamma$ and $x_z$ parametrize the potential
deviations in the electromagnetic and weak dipole couplings from the SM
predictions.

 We see that, for $\gamma=0$, there is only one parameter different from zero,
$\delta_z$, whereas for $k=0$ (that is for $b=(g/\gs)^2 (1/\c^2)$
see eq. (6.3)) we have again only one free parameter, say $x_\gamma$, and
the following relations hold
\bea
\delta_z&=&\f{x_\gamma}{2 \s\c}\\
x_z&=&-\f{\s}{\c} x_\gamma
\eea
So, in the general case with $k$ and $\gamma$ different from zero,
BESS is a combination of the previous situations, and eq. (6.10) always holds.

\resection{Discussion}

  The analysis presented here shows that low energy experiments do not put
really stringent bounds in the plane $(b,g/\gs)$ of the BESS parameter space.
However, when there will be the possibility to add to the analysis the
experiments aimed to measure the anomalous couplings (as in the $e^+e^-$
colliders), we will be able to fix the set
 of parameters
$(b,g/\gs,\gamma)$, in terms of $\epsilon_3$, and of the anomalous couplings
$x_\gamma$, $\delta_z$ and $x_z$. In this situation the relation (6.10) would
represent a test of the model. Unfortunately it is impossible to use in the
present context the quantitative bounds for the anomalous couplings
given in the literature (see, for instance ref. \cite{bil}).
In fact it is generally
assumed that $Z$ and $W$ have standard couplings to the fermions, which
is not the case here, because the effective couplings are modified by terms
proportional to $\epsilon_3$. A precise analysis must take into account this
effect in a proper way. However if we assume that there are no cancellations
among the various contributions we can still make a qualitative analysis.
First of all we recall that the parameter $\gamma$ is naturally expected to be
of order $1/\gs$, because we want that the $V$ bosons decouple in the limit
$\gs\to\infty$.
Then let us consider two possibilities.\hfill\break
\noindent
$i)$ In $\epsilon_3$ there is no
accidental cancellation between the $b$ term and
$(g/\gs)^2$. In this case both quantities should be of order
${(\epsilon_3)}_{exp.}-{(\epsilon_3)}_{rad.corr.}\approx 10^{-3}$
and the same would
be for $k$ defined in eq. (6.3). As far as $\gamma$ is concerned we can write
$\gamma=\tilde\gamma/\gs$
and assume $\tilde\gamma\approx {\cal O}(1)$. From the previous estimate we get
$(g/\gs)\approx 3\%$ and therefore $\gamma g^2/4\approx 5\times 10^{-3}$.
In this case we get both $\delta_z$ and $x_\gamma\approx 5\times 10^{-3}$.
This means that, by considering the results of \cite{bil} for the case of a
two parameters model, even a $1~TeV$ $e^+e^-$ collider is not enough to test
these trilinear anomalous couplings.\hfill\break
\noindent
$ii)$ There is an accidental cancellation in $\epsilon_3$,
$b/2=(g/\gs)^2$. In this
case
\be
k=\f{1}{2 c_\theta^2}\Big(\f{g}{\gs}\Big)^2
\ee
This time there are no low-energy restrictions on $\gs$. However we are working
in the limit $\gs>>g$,
which in practice means that $g/\gs$ should be something less
than, say, $10\%$, and the contribution of this term to $\delta_z$ would be
less that $1\%$. Analogously the $\gamma$-contribution is of the same
order giving $x_\gamma\approx 1.5\%$. From ref. \cite{bil} one can see that an
$e^+e^-$ collider somewhat in
between 500 $GeV$ and 1 $TeV$ would be enough for testing these couplings.

A much more complete analysis of the restrictions of the BESS parameters
on the plane $(b,g/\gs)$, for $\gamma=0$, can be found in ref. \cite{ee}.

 \begin{center}
  \begin{Large}
  \begin{bf}
  Figure Captions
  \end{bf}
  \end{Large}
  \end{center}
\begin{description}
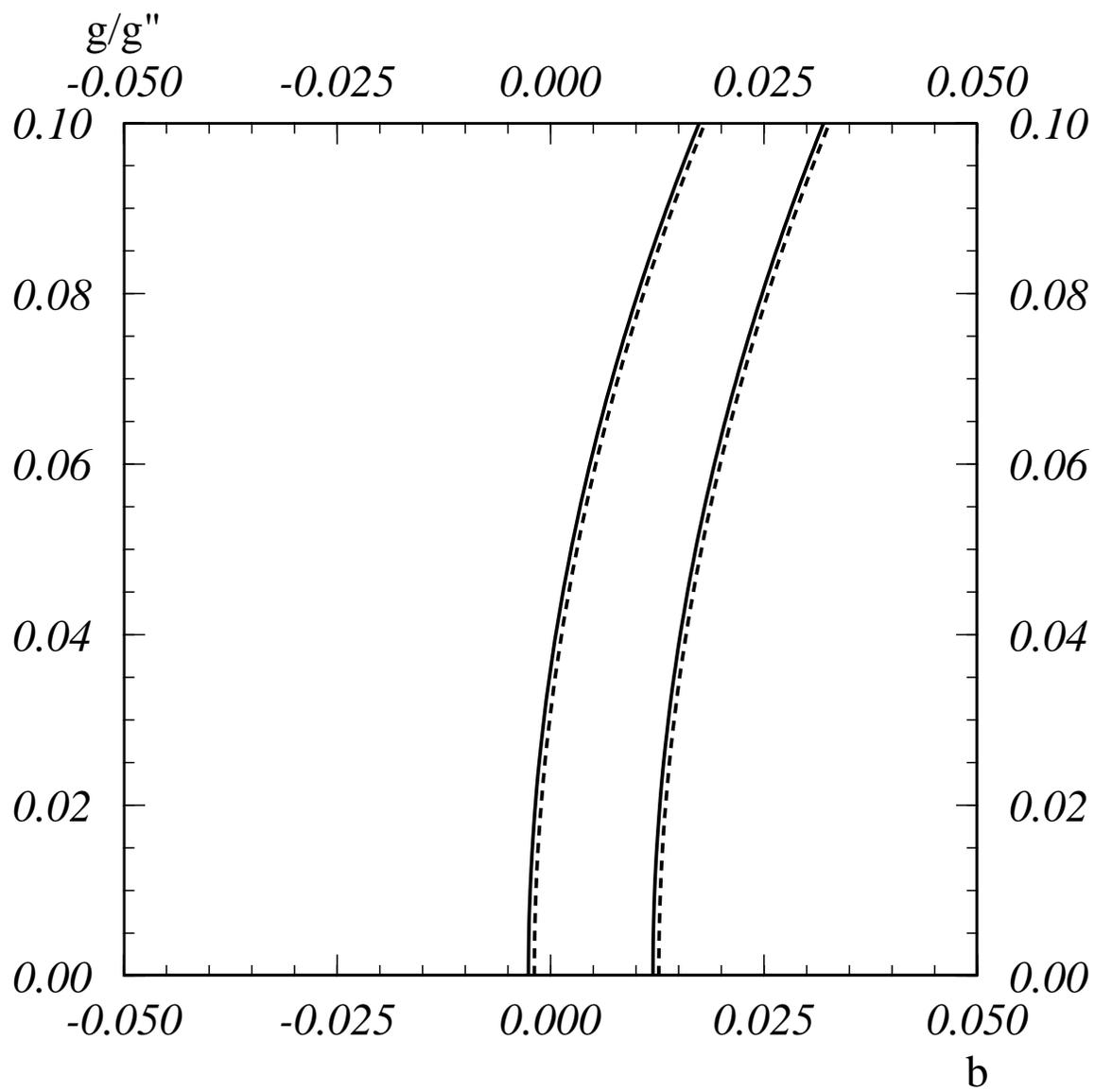
\item [Fig. 1]
Allowed region at $90\%$ C.L. in the plane $(b,g/\gs)$  for the BESS model
coming from the measurement of $\eps_3$.
The solid (dashed) line is for $m_t=191(157)~GeV$ and $\Lambda=1~TeV$.
\end{description}
\end{document}